\documentclass[onecolumn,floatfix,prd,aps,10pt]{revtex4-2}
\usepackage[applemac]{inputenc}
\usepackage[T1]{fontenc}
\pdfoutput=1
\usepackage{enumerate}
\usepackage{braket}
\usepackage{amsmath}
\usepackage{subfigure}
\usepackage{lipsum}
\usepackage{amsfonts}
\usepackage{amssymb, mathrsfs}
\usepackage{bm}
\usepackage[pdftex]{graphicx}
\usepackage{footnote}
\usepackage{hyperref}  
\hypersetup{colorlinks=true,allcolors=blue}
\usepackage{hypcap}   
\usepackage{bookmark}
\usepackage{dsfont,mathrsfs,color,url,verbatim,booktabs}

\def\beq{\begin{equation}}
	\def\eeq{\end{equation}}
\def\bsp{\begin{split}}
	\def\esp{\end{split}}
\def\bea{\begin{eqnarray}}
	\def\eea{\end{eqnarray}}
\def\ba{\begin{array}}
	\def\ea{\end{array}}

\def\l.{\left.}
\def\r.{\right.}

\def\part{\partial}
\def\tfrac#1#2{{\textstyle{#1\over #2}}}

\begin{document}

\title{Electromagnetic Kantowski--Sachs Solutions in Teleparallel $F(T)$ Gravity}
\author{A. Landry}
\email{a.landry@dal.ca}
\affiliation{Department of Mathematics and Statistics, Dalhousie University, Halifax, Nova Scotia, Canada, B3H 3J5}

\begin{abstract}


		A covariant reconstruction framework for electromagnetic Kantowski--Sachs (KS) geometries in teleparallel $F(T)$ gravity is developed using the coframe/spin-connection (CSC) formalism and the invariant approach. In a restricted Maxwell-compatible branch, the electromagnetic conservation laws strongly constrain the anisotropic KS scale factors and lead to the scaling $\rho_{\mathrm{em}}\propto A_3^{-4}$. The corresponding symmetric and antisymmetric field equations are derived and used to reconstruct the functional form of $F(T)$ directly from the KS dynamics. Power-law and exponential ans\"atze generate distinct invariant reconstruction branches associated with electric, magnetic, and transverse electromagnetic sectors. The exponential branch naturally admits reduced teleparallel de Sitter limits and shifted models of the form $F(T)=f(T_0-T)$. The reconstructed branches describe anisotropic cosmological sectors
together with local BH-interior-like sectors that may reproduce reduced BH-interior-like or RN--dS-type behaviors at the level
of the KS dynamics. These branches are organized through the invariant coframe/spin-connection classification and screened using the necessary leading-order viability conditions $F_T>0$ and $F_{TT}>0$. The local and branch-dependent nature of the construction is emphasized throughout.


\end{abstract}


\maketitle


\section{Introduction}\label{sect1}

Teleparallel $F(T)$ gravity provides a conceptually distinct geometric formulation of gravitation in which the gravitational interaction is encoded in spacetime torsion rather than curvature \cite{Aldrovandi2013,Krssak2016,Cai2016,Bahamonde2021}. In this framework, gravity can be interpreted as a gauge theory of translations, described by a coframe $h^{a}{}_{\mu}$ and a spin-connection $\omega^{a}{}_{b\mu}$ defining a curvature-free but torsion-full geometry \cite{Hehl1995,Obukhov2006,Hohmann2018,KrssakPereira2015}. This viewpoint places gravitation on a similar conceptual footing as other fundamental interactions, notably electromagnetism, which is governed by gauge principles.

{Unlike curvature-based approaches in general relativity (GR), covariant teleparallel gravity provides a natural framework in which anisotropic electromagnetic sectors can directly constrain the torsional structure of spacetime through the coframe/spin-connection (CSC) pair \cite{Krssak2016,Golovnev2017,Krssak2019,ColeyLandry2024,nonvacSSpaper,Landry2024,roberthudsonSSpaper,TdSpaper}. In particular, Kantowski--Sachs (KS) geometries constitute a useful testing ground for studying how Maxwell conservation laws restrict admissible anisotropic torsional branches and thereby guide the reconstruction of nontrivial $F(T)$ sectors. This differs from many standard reconstruction approaches in modified gravity, where the functional form of the gravitational action is imposed \emph{{a priori} 
}.}

A major development in modern teleparallel gravity is the fully covariant formulation based on independent CSC pairs, ensuring local Lorentz invariance and a consistent separation between inertial and gravitational effects \cite{Krssak2016,Golovnev2017,Krssak2019}. This formulation resolves long-standing ambiguities related to frame dependence and provides a robust geometric foundation for modified teleparallel theories such as $F(T)$ gravity. In this context, the choice of coframe becomes an intrinsic geometric ingredient rather than a mere computational~tool.

{These considerations naturally motivate the use of invariant classification methods. In particular, the teleparallel invariant program extends the Cartan--Karlhede (CK) algorithm to teleparallel geometries, allowing a systematic classification of spacetimes at the level of the coframe and spin-connection \cite{Coley2009,Coley2011,Krssak2019,Coley:2019zld,ColeyLandry2024,nonvacSSpaper,Landry2024,roberthudsonSSpaper,TdSpaper,McNutt2023}. This approach encodes spacetime symmetries through Lie-derivative conditions and provides a systematic framework for {identifying physically inequivalent reconstruction branches beyond purely metric-based analyses  \cite{ColeyLandry2024,Landry2024,roberthudsonSSpaper,McNutt2023}.}}

Among anisotropic cosmological models, the Kantowski--Sachs (KS) spacetime plays a central role. Originally introduced in \cite{Kantowski1966}, KS geometries describe homogeneous but anisotropic universes with a four-dimensional isometry group, and arise naturally both in early-universe cosmology and in the interior region of Schwarzschild black holes \mbox{(BHs) \cite{Ryan1975,Ellis1969,HawkingEllis}.} Their dynamical properties have been extensively studied in general relativity (GR) and modified gravity theories, including $f(R)$, $f(T)$, $f(Q)$ and related \mbox{frameworks \cite{Boehmer2011,Nashed2021,Rodrigues2015,Leon2014,Dimakis2023,Millano2024}.} In teleparallel gravity, KS spacetimes provide a minimal yet non-trivial setting where torsion dynamics can differ from the corresponding GR description while remaining analytically tractable \cite{Landry2024,Amir2015}. {Several KS solutions have previously been studied in TEGR and nonlinear teleparallel $F(T)$ gravity, motivating the search for covariant reconstruction schemes in anisotropic settings.

	Previous studies of KS or related anisotropic cosmologies in modified gravity have mainly focused on perfect fluids, scalar fields, or predefined teleparallel actions \cite{Landry2024,nonvacSSpaper}. In many cases, the gravitational sector is fixed before solving the field equations (FEs), and the electromagnetic contribution is either neglected or modeled through effective isotropic fluids. By contrast, the present work considers a covariant electromagnetic sector satisfying the Maxwell equations together with the antisymmetric teleparallel FEs (AFEs). Thus, electromagnetic conservation laws act as dynamical selection rules for admissible reconstruction branches. The resulting framework combines covariant teleparallel geometry, invariant classification, and electromagnetic reconstruction in a unified KS framework.

	Electromagnetic fields provide a genuinely anisotropic source sector,
	with stress-energy \mbox{{tensor} 
 \cite{Amir2015,Landryelectro2025}}
	\begin{equation}\label{ElectroEnerMom}
		\Theta^{a}{}_{b} = \mathrm{Diag}(\rho_{\mathrm{em}}, -\rho_{\mathrm{em}}, \rho_{\mathrm{em}}, \rho_{\mathrm{em}}),
	\end{equation}
	where $\rho_{em}=-P_r=P_t$ in the orthonormal KS coframe. This structure naturally matches the
	anisotropy of KS geometries and is relevant both for anisotropic
	cosmological models and for BH-interior-like regimes. Electromagnetic fields may also affect the causal structure and stability properties of strong-gravity KS regimes. In teleparallel gravity, such effects are further modified by the torsional response to the Maxwell field \cite{Cai2016,Krssak2019}.}

Within the restricted Maxwell-compatible branch considered below, the
Maxwell conservation laws impose a characteristic scaling relation between
the electromagnetic energy density and the angular KS scale factor. This
scaling provides the source constraint used in the reconstruction procedure \cite{Jackson1999,LandauLifshitz1975,HehlObukhov2003}.

 {
	
 The aim of this work is therefore to develop} a unified and systematic framework for constructing and organizing electromagnetic KS reconstruction branches in covariant $F(T)$ gravity. Our approach combines three key ingredients: (i) the covariant CSC formalism, (ii)~the invariant classification, and {(iii) the Maxwell equations and associated conservation constraints.} Unlike many previous studies relying on predefined $F(T)$ ansatze, we adopt a reconstruction strategy in which the functional form of $F(T)$ is determined directly from the FEs. {The main novelties of the present work are: (i) a covariant CSC treatment
 of electromagnetic KS geometries, (ii) the use of Maxwell conservation
 laws as reconstruction constraints, and (iii) the organization of the
 resulting power-law, exponential and teleparallel de Sitter branches
 within the invariant framework.} Throughout the paper, KS geometries are interpreted in two complementary ways: as homogeneous anisotropic cosmologies and as effective descriptions of BH-interior-like sectors. {This dual interpretation motivates the analysis of both anisotropic cosmological expansion and local KS-interior-like reconstruction sectors.

The reconstruction procedure developed in this work is local and branch-dependent. The resulting KS branches should therefore be interpreted as invariant reconstruction sectors associated with specific CSC branches rather than as complete global compact-object geometries. In particular, the BH-interior-like and RN--dS-like sectors identified below correspond, at the level of reduced KS dynamics, to local KS reduced branches whose global extension, horizon structure, and perturbative stability require additional analysis. Similarly, the stability conditions $F_T>0$ and $F_{TT}>0$ used throughout the paper should be understood as necessary leading-order viability conditions rather than as a complete perturbative treatment of anisotropic KS backgrounds.

The main objectives of this work are the following:
\begin{itemize}
	\item Derive the covariant symmetric and antisymmetric $F(T)$ FEs (SFEs and AFEs) for electromagnetic KS geometries within the CSC formalism;
	\item Determine the restrictions imposed by the Maxwell conservation laws on admissible anisotropic KS branches;
	\item Reconstruct the teleparallel action $F(T)$ from invariant KS dynamics using power-law and exponential ans\"atze;
	\item Organize the resulting electric, magnetic, and transverse electromagnetic reconstruction sectors within the invariant framework;
	\item Screen the associated cosmological, teleparallel de Sitter, and local KS-interior-like branches using leading-order viability conditions.
\end{itemize}

The paper is organized as follows. Section \ref{sect2} presents the covariant teleparallel FEs, the electromagnetic sector, and the KS CSC geometry. Sections \ref{sect3} and \ref{sect4} develop the power-law and exponential reconstruction branches, respectively. Section \ref{sect5} extends the analysis to more general effective electromagnetic sectors and local KS-interior-like reconstruction branches. Finally, Section \ref{sect6} summarizes the main results and discusses their physical implications and limitations.

}


\section{Teleparallel Field Equations, Maxwell Sector, and Kantowski--\linebreak  Sachs Geometry}\label{sect2}

{
The purpose of this section is to specify the geometric branch, matter
sector, and reduced field equations used in the reconstruction analysis.
All subsequent branches are constructed within this fixed covariant
coframe--spin-connection setting.
}

\subsection{Teleparallel $F(T)$ Gravity and General Conservation Laws}

In the covariant formulation of teleparallel gravity, the gravitational dynamics is described by a generalized $F(T)$ action including a minimally coupled electromagnetic sector. The action is given by \cite{Cai2016, Krssak2019, Aldrovandi2013,ColeyLandry2024,Landry2024}:
\begin{equation}
	S_{F(T)} = \int d^4 x \left[ \frac{h}{2\kappa} F(T) - \frac{1}{4} F_{\mu\nu} F^{\mu\nu} \right],
	\label{action}
\end{equation}
{where $h = \det(h^a_{\ \mu})$ is the coframe determinant, $T$ is the torsion scalar, and \mbox{$F_{\mu\nu} = \nabla_\mu A_\nu - \nabla_\nu A_\mu$} is the electromagnetic field tensor.}

Varying the action with respect to the coframe $h^a_{\ \mu}$ yields the general FEs in covariant $F(T)$ gravity \cite{Krssak2016,Hohmann2018}:
\begin{align}
	\kappa \Theta^\mu_{\ a} &=
	h^{-1} F_T \partial_\nu (h S^{\mu\nu}_{\ \ a})
	+ F_{TT} S^{\mu\nu}_{\ \ a} \partial_\mu T
	+ \frac{F}{2} h^\mu_{\ a} \nonumber \\
	&\quad - F_T T^b_{\ a\nu} S^{\mu\nu}_{\ \ b}
	- F_T \omega^b_{\ a\nu} S^{\mu\nu}_{\ \ b},
	\label{FE_full}
\end{align}
where $F_T = dF/dT$ and $F_{TT} = d^2F/dT^2$, $S^{\mu\nu}_{\ \ a}$ is the superpotential, $T^b_{\ a\nu}$ is the torsion tensor, and $\omega^b_{\ a\nu}$ is the spin-connection. 

These equations can be decomposed into symmetric and antisymmetric parts \cite{Krssak2019, Hohmann2018}:
\begin{equation}
	\kappa \Theta_{(ab)} = F_T \overset{\circ}{G}_{ab}
	+ F_{TT} S^\mu_{\ (ab)} \partial_\mu T
	+ \frac{g_{ab}}{2} \left[ F - T F_T \right],
	\label{FE_sym}
\end{equation}
\begin{equation}
	0 = F_{TT} S^\mu_{\ [ab]} \partial_\mu T.
	\label{FE_antisym}
\end{equation}

{The AFEs represent purely teleparallel consistency conditions with no counterpart in GR.} In particular, nonlinear \(F(T)\) branches with \(F_{TT}\neq0\) require compatibility between \(\partial_t T\) and the antisymmetric superpotential sector. {The TEGR limit corresponds to $F(T)=T+\textrm{const.}$,	for which $F_{TT}=0$ and Equation~\eqref{FE_antisym} becomes identically satisfied.}

From the matter sector, the canonical energy--momentum tensor is defined through the variation of the matter Lagrangian $\mathcal{L}_{\text{Source}}$ as \cite{Hehl1995, Obukhov2006}:
\begin{equation}
	\Theta^\mu_{\ a} = \frac{1}{h} \frac{\delta \mathcal{L}_{\text{Source}}}{\delta h^a_{\ \mu}},
	\label{EM_tensor}
\end{equation}
and satisfies the covariant CL:
\begin{equation}
	\overset{\circ}{\nabla}_\nu \Theta^{\mu\nu} = 0.
	\label{conservation}
\end{equation}

{{For minimally} 
coupled Maxwell fields, Equation~\eqref{conservation} follows directly from the diffeomorphism invariance of the matter action and does not depend on the specific functional form of $F(T)$. Equation~\eqref{conservation}} corresponds to the standard conservation of energy--momentum in GR, now arising from diffeomorphism invariance within the teleparallel framework. Decomposing into symmetric and antisymmetric components, one obtains
\begin{equation}
	\Theta_{[ab]} = 0, \qquad \Theta_{(ab)} = T_{ab},
	\label{symmetry}
\end{equation}
where $T_{ab}$ denotes the physical symmetric energy--momentum tensor.

This structure assumes minimal coupling between matter and the metric, implying a vanishing hypermomentum. In the general case, a non-zero hypermomentum leads to modified CLs. From Equations~(\ref{FE_sym}) and (\ref{FE_antisym}), one can rewrite the SFEs as an effective gravitational balance relation \cite{Obukhov2006, Krssak2019}:
\begin{equation}
	0 = \kappa \Theta_{ab}
	- F_T \overset{\circ}{G}_{ab}
	- F_{TT} S^\mu_{\ ab} \partial_\mu T
	- \frac{g_{ab}}{2} \left[ F - T F_T \right] .
	\label{hypermomentum}
\end{equation}

{Equation}~\eqref{hypermomentum} is therefore not an additional CL but rather a convenient rewriting of the SFEs in terms of an effective gravitational balance. It illustrates that, in teleparallel gravity, the effective balance equations are directly linked to the torsional structure of spacetime and to the functional form of \(F(T)\). For minimally coupled electromagnetic matter with vanishing hypermomentum, the usual metric-compatible CL in Equation~\eqref{conservation} is recovered, ensuring consistency with GR solutions.

 {This formulation} 
 emphasizes the dynamical origin of the CLs within the covariant teleparallel framework. In teleparallel $F(T)$ gravity, the presence of \(F_T\) and \(F_{TT}\) {terms modifies the effective gravitational response to matter through the torsional sector.}

\subsection{Teleparallel Kantowski--Sachs Coframe/Spin-Connection Pair from Lie Algebra}\label{sect23}

In the covariant formulation of teleparallel gravity, admissible geometries are determined by symmetry requirements imposed simultaneously on the coframe and the spin-connection. These requirements are naturally expressed through Lie derivatives along the Killing vectors (KVs) generating the spacetime isometry group. A teleparallel geometry must satisfy \cite{Coley2009,Coley2011,Coley:2019zld,Krssak2019}:
\begin{equation}
	\mathcal{L}_X h^a = \lambda^a_{\ b} h^b, 
	\qquad 
	\mathcal{L}_X \omega^a_{\ bc} = 0,
	\label{11}
\end{equation}
where $h^a$ is the orthonormal coframe, $\mathcal{L}_X$ denotes the Lie derivative along a KV $X$, and $\lambda^a_{\ b}$ generates local Lorentz transformations $\Lambda^a_{\ b}$ {belonging to the Lorentz algebra $\mathfrak{so}(1,3)$.}

In addition, a proper teleparallel geometry must satisfy the vanishing curvature condition, ensuring that gravitation is entirely encoded in torsion rather than \mbox{curvature \cite{Krssak2016,Krssak2019,Hehl1995}:}
\begin{equation}
	R^a_{\ b\mu\nu}
	= \partial_\mu \omega^a_{\ b\nu}
	- \partial_\nu \omega^a_{\ b\mu}
	+ \omega^a_{\ e\mu} \omega^e_{\ b\nu}
	- \omega^a_{\ e\nu} \omega^e_{\ b\mu}
	= 0
	\;\;\Rightarrow\;\;
	\omega^a_{\ b\mu} = \Lambda^a_{\ c} \partial_\mu \Lambda^c_{\ b}.
	\label{12}
\end{equation}

A natural orthonormal coframe adapted to time-dependent KS geometries in coordinates $(t,r,\theta,\phi)$ is given by \cite{ColeyLandry2024,Landry2024,roberthudsonSSpaper,McNutt2023}:
\begin{equation}
	h^a_{\ \mu} = \mathrm{Diag}\left[1,\; A_2(t),\; A_3(t),\; A_3(t)\sin\theta \right].
	\label{13}
\end{equation}

{{The diagonal} form \eqref{13} is compatible with the	homogeneous KS isometry group and provides the simplest representative within the corresponding teleparallel equivalence class.} Without loss of generality, we fix $A_1(t)=1$ through a time reparametrization, a standard procedure in homogeneous cosmologies \cite{Ryan1975,Ellis1969}.

An alternative parametrization is obtained by imposing \(A_3(t)=t\):
\begin{equation}
	h^a_{\ \mu} = \mathrm{Diag}\left[A_1(t),\; A_2(t),\; t,\; t\sin\theta \right].
	\label{14}
\end{equation}

{{This branch} is useful for comparison with areal-time descriptions commonly employed in BH-interior analyses.} In the remainder of Sections~\ref{sect3}--\ref{sect5}, however, we use the cosmological-time gauge \(A_1=1\), which is better suited to reconstruction in homogeneous \mbox{cosmology \cite{ColeyLandry2024,Landry2024,roberthudsonSSpaper,McNutt2023}.}

Although Equations~\eqref{13} and \eqref{14} can be rewritten using tetrads $e^a_{\ \mu}$ in orthonormal gauges, the coframe formulation $h^a_{\ \mu}$ is preferable in covariant teleparallel gravity. Indeed, the theory is invariant under local Lorentz transformations, and the CSC pair provides a consistent separation between inertial and gravitational effects \cite{Krssak2016,Krssak2019,Hohmann2018}. 

In contrast, tetrad-only formulations may lead to frame-dependent results or ambiguities in the presence of non-trivial spin-connections \cite{Golovnev2017,Krssak2016}. The covariant approach resolves these issues and ensures the physical consistency of solutions.

{The non-vanishing components of the spin-connection $\omega^a_{\ bc}$ compatible with KS symmetry can be parametrized as $\omega^a_{\ bc} = \omega^a_{\ bc}(\psi(t), \chi(t))$, where $\psi$ and $\chi$ are arbitrary \mbox{functions \cite{ColeyLandry2024,Landry2024,roberthudsonSSpaper,McNutt2023}.} }

Imposing the $F(T)$ AFEs (cf. Equation~\eqref{FE_antisym}) yields the constraints
\[
\psi = 0, 
\qquad 
\chi = \frac{\pi}{2}
\quad (\text{or } \tfrac{3\pi}{2}),
\]
which correspond to proper covariant teleparallel frames \cite{Krssak2016,Krssak2019}.

Having fixed the KS coframe branch, we now specify the spin-connection required for covariance and local Lorentz invariance. The resulting non-zero components of the spin-connection are
\begin{equation}
	\omega^2_{\ 34} = -\omega^2_{\ 43} = \delta, 	\qquad 	\omega^3_{\ 44} = -\frac{\cos\theta}{A_3 \sin\theta},
	\label{15}
\end{equation}
with $\delta = \pm 1$ corresponding to locally equivalent discrete teleparallel branches  {that generate the same metric geometry but differ by a discrete
teleparallel gauge choice \cite{ColeyLandry2024,Landry2024}. This spin-connection coincides with the proper inertial connection required to remove spurious inertial torsion contributions and restore local Lorentz covariance.}

The coframe--spin-connection pair defined by Equations~\eqref{13}--\eqref{15} satisfies the symmetry conditions, the zero-curvature condition, and the AFEs, thereby defining a fully consistent covariant teleparallel KS geometry.

Moreover, this construction is equivalent to that obtained via the CK invariant classification extended to teleparallel gravity \cite{Coley2009,Coley2011,Coley:2019zld,McNutt2023}. This approach classifies spacetimes at the level of the coframe and spin-connection rather than solely through the metric, allowing for the identification of physically inequivalent geometries that would otherwise appear identical in purely metric-based analyses. This illustrates how the invariant approach provides a systematic organization of teleparallel
reconstruction branches \cite{ColeyLandry2024,Landry2024}.

\subsection{Electromagnetic Source Conservation~Laws}\label{sect24}

For the KS spacetime defined by Equation~\eqref{13}, the electromagnetic CLs are
\begin{align}
	0=& \partial_t\rho_{em}+4\partial_t\left(\ln\,A_3(t)\right)\,\rho_{em},  \label{KSCLs}
\end{align}
where $\rho_{em}=\tfrac{1}{2}\left(E^2+B^2\right)$. The \(E(t)\) and \(B(t)\) terms denote the physical radial electric and magnetic amplitudes measured in the orthonormal coframe. We simplify Equation~\eqref{KSCLs} as
\begin{align}
	0=& \partial_t\,\ln\left[\rho_{em}(t)\,A_3^4(t)\right]\quad \Rightarrow\quad \rho_{em}(t)=\frac{\rho_{em\,0}}{A_3^4(t)}.  \label{KSCLsolution}
\end{align}

{{In} the present work, Equation~\eqref{KSCLsolution} is adopted as the defining invariant electromagnetic scaling of the reconstruction branch. The Maxwell solutions are then used to identify the subclass of
 	electromagnetic configurations compatible with this scaling. Equation~\eqref{KSCLsolution} should be understood as a consequence of the KS symmetry assumptions and of the restricted Maxwell-compatible branch considered throughout this work. Hereafter, the term restricted Maxwell-compatible branch refers to the subclass of Maxwell solutions satisfying Equation~\eqref{KSCLsolution}. More general Maxwell branches may lead to different scaling~laws.}

The electromagnetic sector must also satisfy the covariant Maxwell equations:
\begin{align}\label{1001maxwell}
	\nabla_{\mu}F^{\mu\nu}=J^{\nu}\quad \text{and}\quad  \nabla_{\alpha}F_{\mu\nu}+\nabla_{\nu}F_{\alpha\mu}+\nabla_{\mu}F_{\nu\alpha}= 0 .
\end{align}

{These} constraints distinguish the time-dependent KS electromagnetic sector from the static SS electromagnetic solutions studied previously \cite{ColeyLandry2024,nonvacSSpaper,Jackson1999,LandauLifshitz1975,HehlObukhov2003}. We add the conserved $4$-current:
\begin{equation}\label{electriccurrent}
	\nabla_{\nu}J^{\nu}=0,
\end{equation}
{for a completely conserved electromagnetic solution. After projecting the Maxwell equations onto the orthonormal KS coframe, the independent electric, magnetic, and current components reduce to Equation~\eqref{Maxwellfield}. For the KS spacetime, Equations \eqref{1001maxwell} and \eqref{electriccurrent} reduce to}

\begin{align}
	A_2A_3^2\rho_{\rm elec}
	&=
	\partial_t(A_2A_3^2E),
	\quad
	A_3^2J^r
	=
	\partial_t(A_3^2E),
	\nonumber\\
	0&=\partial_t(A_3^2B),
	\quad
	0=\partial_t(A_2A_3^2\rho_{\rm elec}), \label{Maxwellfield}
	\\
	\rho_{elec}(t)=&\frac{\rho_{elec\,0}}{A_2(t)\,A_3^2(t)},\quad E(t)= \frac{\left(\rho_{elec\,0}\,t+E_0\right)}{A_2(t)\,A_3^2(t)} \quad B(t)=\frac{B_0}{A_3^2(t)}
	\nonumber\\
	J^r(t)=&\frac{1}{A_2^2(t)\,A_3^2(t)}\left[{\rho_{elec\,0}\left(A_2(t)-t\partial_tA_2(t)\right)-E_0\partial_tA_2(t)}\right].  \label{electrosol}
\end{align}

{{The} linear dependence on $t$ follows from the time integration of the Maxwell conservation equations in the cosmological-time gauge $A_1=1$.} The product \(A_2(t)A_3^2(t)\) is proportional to the comoving spatial volume element of the KS geometry. Substituting Equation~\eqref{electrosol} into Equation~\eqref{KSCLsolution} yields
\begin{equation}\label{CLconstrained}
	\rho_{em}(t)=\frac{1}{2A_3^4(t)}\left(\frac{\left(\rho_{elec\,0}\,t+E_0\right)^2}{A_2^2(t)}+B_0^2\right)=\frac{\rho_{em\,0}}{A_3^4(t)}.
\end{equation}

{{Equation}~\eqref{CLconstrained} shows explicitly that the scaling \eqref{KSCLsolution} is not automatically satisfied by the general Maxwell solution.} If one imposes the pure angular flux scaling \(\rho_{\rm em}=\rho_{{\rm em},0}A_3^{-4}\) together with the Maxwell solutions \eqref{electrosol}, then Equation~\eqref{CLconstrained} defines a restricted Maxwell-compatible branch. In this branch, one obtains
\begin{equation}\label{eqn23}
	A_2(t)
	=
	\pm
	\frac{\rho_{{\rm elec},0}t+E_0}
	{\sqrt{2\rho_{{\rm em},0}-B_0^2}}
	=
	A_{20}(t+\tilde E_0),
\end{equation}
where 
\begin{equation}
	A_{20}
	=
	\pm
	\frac{\rho_{{\rm elec},0}}
	{\sqrt{2\rho_{{\rm em},0}-B_0^2}},
	\qquad
	\tilde E_0=\frac{E_0}{\rho_{{\rm elec},0}},
	\qquad
	2\rho_{{\rm em},0}>B_0^2,
\end{equation}
with $\rho_{{\rm elec},0} \neq 0$. The condition \(2\rho_{{\rm em},0}>B_0^2\) ensures that the restricted branch remains real-valued. This branch corresponds to a special Maxwell-compatible reconstruction branch associated with the pure angular flux scaling. It should not be interpreted as the most general Maxwell solution in KS geometry. More general current profiles may relax this constraint and lead to different \(A_2(t)\) evolutions.

{Using the ansatz formulation in Equation~\eqref{14}, Equation~\eqref{KSCLs} becomes} 
\begin{equation}
	0= \partial_t\rho_{em}+\frac{4}{t}\rho_{em}, \quad\Rightarrow\quad \rho_{em}=\frac{\rho_{em\,0}}{t^4} . \label{KSCLs2}
\end{equation} 

{{The} solutions of Equations~\eqref{Maxwellfield} and \eqref{electrosol} are summarized as}
\begin{align}\label{eqn26}
	\rho_{elec}(t)=&\frac{\rho_{elec\,0}}{A_2(t)\,t^2}\quad	E(t)=\frac{1}{A_2(t)\,t^2}\left[\rho_{elec\,0}\int\,dt'\,A_1(t')+E_0\right],	\quad	B(t)=\frac{B_0}{t^2}, \quad 
	\nonumber\\
	J^r(t)=&
	\frac{1}{A_1A_2^2t^2}
	\frac{d}{dt}
	\left[
	\frac{\rho_{{\rm elec},0}\int^t A_1(t')dt' + E_0}{A_2(t)}
	\right].
\end{align}	

{In} the \(A_3=t\) branch, the current component \(J^r(t)\) generally does not vanish unless additional constraints are imposed on \(A_1(t)\), \(A_2(t)\), and the electric source constants. In the remainder of the paper, we mainly use the \(A_1=1\) cosmological-time branch, while Equation~\eqref{eqn26} is retained for comparison with the areal-time parametrization.


\subsection{Symmetric and Unified Field~Equations}\label{sect25}

With the KS-compatible CSC pair fixed, the torsion scalar and the independent SFEs can be written entirely in terms of the two scale factors \(A_2(t)\) and \(A_3(t)\). These equations form the dynamical core of the reconstruction procedure. The torsion scalar and the SFE components for the $\chi = \frac{\pi}{2}$ ($\delta=+1$) case in Equation~\eqref{FE_sym} are~\cite{ColeyLandry2024,Landry2024}:
\begin{align}
	T =& 2\left(\ln(A_3)\right)'\left(\left(\ln(A_3)\right)'+2\left(\ln(A_2)\right)'\right) - \frac{2}{A_3^2} , \label{301a}
	\\
	\kappa \rho_{em} =& -\tfrac{1}{2}\left[F-TF_T\right]+F_T\left[\frac{1}{A_3^2}+2\left(\ln\,A_2\right)'\left(\ln\,A_3\right)'+\left(\ln\,A_3\right)'^2\right] , \label{301b} 
	\\
	-\kappa \rho_{em} =& \tfrac{1}{2}\left[F-TF_T\right]-2\partial_t\left(F_T\right)\left(\ln\,A_3\right)'-F_T\left[\frac{1}{A_3^2}+\frac{2A_3''}{A_3}+\left(\ln\,A_3\right)'^2\right] , \label{301c}
	\end{align}	 
	
	\vspace{-12pt}
	
\begin{align}
	\kappa \rho_{em} =& \tfrac{1}{2}\left[F-TF_T\right]-\partial_t\left(F_T\right)\left[\left(\ln\,A_3\right)'+\left(\ln\,A_2\right)'\right]-F_T\left[\frac{A_2''}{A_2}+\frac{A_3''}{A_3}+\left(\ln\,A_2\right)'\left(\ln\,A_3\right)'\right]  , \label{301d}
\end{align}
\noindent where \(F_T\) is not assumed to be constant. {In Equation~\eqref{301a}, the first term corresponds to anisotropic expansion contributions, whereas the second term originates from the angular KS sector.} Compared with some KS \(F(T)\)-gravity formulations in the literature~\cite{Rodrigues2015,Amir2015}, the present covariant CSC formulation leads to modified component expressions. For the \(\delta=-1\) branch, only minor sign changes occur in selected terms of Equations~\eqref{301b}--\eqref{301d}, corresponding to sign changes in selected connection-dependent terms. Apart from these sign changes, the general structure of Equations~\eqref{301b}--\eqref{301d} remains unchanged.

Combining Equations~\eqref{301b} and \eqref{301c}, and assuming \(A_3'\neq0\), eliminates the \([F-TF_T]\) and \(\rho_{\rm em}\) terms and yields
\begin{align}
	\partial_t\left(\ln F_T\right) =& \left[\left(\ln\,A_2\right)' -\frac{A_3''}{A_3'}\right] . \label{302}
\end{align}

{{This} equation is independent of the explicit electromagnetic source normalization and therefore governs the invariant reconstruction sector common to all Maxwell branches. Consequently, different Maxwell sectors enter the reconstruction primarily through the source equation, while Equation~\eqref{302} controls the underlying invariant gravitational branch. The same equation should be interpreted as a local invariant reconstruction relation valid on a fixed CSC branch. This relation assumes that $F_T(T)\neq 0$ and that $F_T$ does not change sign on the branch considered; it therefore excludes degenerate TEGR-like branches with constant $F_T$. It plays the role of the reduced invariant reconstruction equation} for the KS sector and represents the KS analogue of the reconstruction equation used in the static SS case, with the radial derivative replaced by the time derivative. Equation~\eqref{302} is valid on branches where \(F_T\neq0\), \(A'_3\neq0\), and \(\partial_t\ln F_T\) is well defined; branches violating these conditions must be treated separately.
Equations~\eqref{301a}, \eqref{301d} and \eqref{302} together provide the master teleparallel FEs system used in Sections~\ref{sect3}--\ref{sect5}.


\section{Power-Law Reconstruction Branches in Kantowski--Sachs\linebreak   Teleparallel Gravity}\label{sect3}

{In this section, we construct local power-law reconstruction branches of the reduced FEs \eqref{301a}--\eqref{301d} using a PL ansatz within the covariant teleparallel framework. The reconstruction is performed on fixed locally invertible CSC branches satisfying the assumptions stated after Equation~\eqref{302}.} We adopt the teleparallel invariant approach based on invariant classification and the CSC pair \cite{ColeyLandry2024,Landry2024,roberthudsonSSpaper}, ensuring the consistency of the teleparallel geometry.

\subsection{Power-Law Ansatz and Torsion Scalar}

We consider the KS coframe ansatz with
\begin{equation}
	A_1 = 1, \quad A_2(t) = b_0 t^b, \quad A_3(t) = c_0 t^c.
\end{equation}

{We} restrict to \(t>0\) so that the PL branches are real and the logarithmic derivatives are well defined. The exponent \(b\) controls the radial scale factor while \(c\) controls the angular two-sphere. Their difference measures the anisotropic shear of the KS geometry.

The logarithmic derivatives are
\begin{equation}
	H_2=\frac{\dot A_2}{A_2}=\frac{b}{t},\qquad 	H_3=\frac{\dot A_3}{A_3}=\frac{c}{t},\qquad 	\sigma^2\propto (H_2-H_3)^2.
\end{equation}

The torsion scalar becomes
\begin{equation}\label{eqn34}
	T = \frac{2c(c+2b)}{t^2} - \frac{2}{c_0^2 t^{2c}}.
\end{equation}

{The} sign of \(T\) depends on the balance between the kinetic torsion contribution and the angular-curvature term. {The first term in Equation~\eqref{eqn34} is controlled by the anisotropic expansion rates, while the second term is inherited from the angular KS sector. Consequently, the sign and monotonicity of \(T(t)\) are branch-dependent and must be checked before using \(T\) as a reconstruction variable.}

Three integrable cases emerge:

\begin{itemize}
	\item $c=1$: $T \sim t^{-2}$ (scaling regime). This case is especially useful because both the angular-curvature term and the kinetic torsion contribution scale as \(t^{-2}\), allowing for a direct algebraic inversion \(t=t(T)\).
	\item $c=-2b$: $T \sim t^{-2c}$ (pure angular--torsional regime). In this branch, the kinetic contribution \(2c(c+2b)t^{-2}\) vanishes and the torsion scalar is controlled entirely by the angular-curvature term.
	\item $c\neq \left\{1,\,-2b\right\}$ (general or intermediate regimes):
	\begin{equation}
		0={2c(c+2b)}\,t^{-2} - \frac{2}{c_0^2}\,t^{-2c}-T.
	\end{equation}
	
\end{itemize}
{Each} branch leads to a specific local inversion \(t=t(T)\), whenever \(\dot T\neq0\), and therefore to a corresponding reconstruction of teleparallel \(F(T)\). The cases \(c=1\) and \(c=-2b\) are the analytically tractable branches used below; the intermediate branch generally requires implicit reconstruction or a case-by-case inversion of \(T(t)\). Most developments and results concern the \(c=1\) branch.

\subsection{Teleparallel Reconstruction Method}

Following \cite{Coley:2019zld,ColeyLandry2024,Landry2024}, we express all physical quantities in terms of the invariant scalar $T$. After substituting the PL ansatz into the reduced system \eqref{301a}, \eqref{301d}, and \eqref{302}, all time-dependent terms can be expressed as powers of \(T\) on locally invertible branches. The resulting equation for \(F(T)\) takes the Euler form
\begin{equation}\label{eqn35}
	T^2F_{TT}+\gamma_1 TF_T+\gamma_0F=\kappa\rho(T),
\end{equation}
where $\gamma_1$ and $\gamma_0$ are branch-dependent constants determined by the KS exponents $(b,c)$ and by the electromagnetic sector under consideration. The coefficients \(\gamma_1\) and \(\gamma_0\) are fixed only after the invariant branch and electromagnetic source scaling have been selected. {The source term $\rho(T)$ is fixed by the effective Maxwell-compatible scaling sector under consideration; hence, the electromagnetic scaling sector determines the particular reconstructed contribution} while the homogeneous part encodes the vacuum teleparallel branch. The homogeneous solution is
\begin{equation}
	F_h(T) = C_1 T^{m_1} + C_2 T^{m_2},
\end{equation}
with
\begin{equation}
	m_{1,2} = \frac{1}{2}\left[1-\gamma_1\pm \sqrt{(\gamma_1-1)^2-4\gamma_0}\right].
\end{equation}

{The} constants \(C_1\) and \(C_2\) parametrize the homogeneous teleparallel sector, while the coefficient \(\lambda\) appearing below is fixed by the electromagnetic source normalization. When the discriminant \((\gamma_1-1)^2-4\gamma_0\) is negative, the homogeneous sector corresponds to logarithmic oscillatory modes in \(T\), which require a separate physical interpretation. The reduction to Equation~\eqref{eqn35} assumes that \(T(t)\) is locally invertible on the branch considered and sufficiently differentiable. Points satisfying \(\dot T=0\) correspond to critical invariant branches and must be treated separately.

{The following electric, magnetic, and transverse sectors should be understood as effective orthonormal electromagnetic scaling sectors. Each sector determines a particular source term in the reduced reconstruction equation, while full Maxwell compatibility must be checked against the constraints discussed in Section \ref{sect24}. }

\subsection{Radial Electric Field}

{For the effective radial electric scaling sector compatible with the reconstruction framework}, the relevant electric density scales as
\[
\rho_E(t)\sim t^{-4b},
\]
which corresponds to the effective orthonormal scaling \(E_r\sim A_2^{-2}\). Here, \(E_r\) denotes the orthonormal radial electric amplitude.

In the scaling case $c=1$, one finds
\begin{equation}
	\rho_E(T) \sim T^{2b}.
\end{equation}

The reconstructed function is therefore
\begin{equation}\label{eqn40}
	F(T) = C_1 T^{m_1} + C_2 T^{m_2} + \lambda T^{2b}.
\end{equation}

{This shows that the radial electric scaling sector carries an explicit dependence on the anisotropy parameter \(b\) in the reconstructed contribution to the gravitational action.}

\subsection{{Radial Magnetic Field} 
}

For a radial magnetic field,
\begin{equation}
	B_r \propto \frac{1}{A_3^2}, \quad \rho_B \sim t^{-4c}.
\end{equation}

For $c=1$, we obtain
\begin{equation}
	\rho_B(T) \sim T^2,
\end{equation}
leading to
\begin{equation}\label{eqn43}
	F(T) = C_1 T^{m_1} + C_2 T^{m_2} + \lambda T^2.
\end{equation}

{This quadratic behavior, within the scaling branch \(c=1\), reflects the way the radial magnetic scaling is mapped into the invariant torsion variable.}

\subsection{{Transverse Electromagnetic Field}}

For transverse fields,
\begin{equation}
	\rho_\perp \sim \frac{1}{A_2^2 A_3^2} \sim t^{-2(b+c)}.
\end{equation}

For $c=1$,
\begin{equation}
	\rho_\perp(T) \sim T^{b+1},
\end{equation}
and thus
\begin{equation}\label{eqn46}
	F(T) = C_1 T^{m_1} + C_2 T^{m_2} + \lambda T^{b+1}.
\end{equation}

{The different powers of \(T\) in Equations~\eqref{eqn40}, \eqref{eqn43}, and \eqref{eqn46} show that different electromagnetic scaling sectors lead to distinct invariant reconstruction powers within the \(c=1\) scaling branch. The corresponding PL branches are summarized in Table~\ref{table1}. All tables below summarize local reconstruction sectors on fixed CSC branches; they should not be interpreted as classifications of complete global spacetime solutions.}

\begin{table}[ht]
		\caption{{Electromagnetic} reconstruction branches in the PL KS sector.}
	\label{table1}
	\begin{tabular}{ccc}
		\toprule
		\textbf{EM Branch} & \textbf{Density Scaling for} \boldmath{\(c=1\)} & \textbf{Reconstructed Term} \\
		\midrule
		Radial electric & \(\rho_E(T)\sim T^{2b}\) & \(T^{2b}\) \\
		Radial magnetic & \(\rho_B(T)\sim T^2\) & \(T^2\) \\
		Transverse & \(\rho_\perp(T)\sim T^{b+1}\) & \(T^{b+1}\) \\
		\bottomrule
	\end{tabular}
\end{table}

Table~\ref{table1} refers to the scaling branch \(c=1\), for which \(T\sim t^{-2}\). Other values of \(c\) lead to different implicit reconstruction branches. The table highlights how {different electromagnetic scaling sectors generate distinct invariant powers} in the reconstructed teleparallel action.

\subsection{Cosmological Solutions}

The scale factors describe anisotropic cosmologies:

\begin{itemize}
	\item \(b=c\): isotropic expansion rates. Strictly speaking, the spatial topology remains KS rather than flat FLRW.
	\item $b>c$: anisotropic shear-dominated phase;
	\item \(c<0\): {contracting angular sector, relevant to KS-interior-like behavior.}
\end{itemize}

The torsion scalar behaves as
\begin{equation}
	T \sim t^{-2} \quad \text{or} \quad T \sim t^{-2c},
\end{equation}
{leading to PL reconstruction branches controlled by the torsion scalar.}

The mean expansion rate and shear are
\[
H=\frac{1}{3}(H_2+2H_3)=\frac{b+2c}{3t},
\qquad
\sigma^2=\frac{1}{3}(H_2-H_3)^2.
\]

{Thus,} the isotropic limit corresponds to \(b=c\), while \(b\neq c\) describes a shear-dominated anisotropic phase. For constant PL exponents \((b,c)\), the dimensionless ratio \(\sigma^2/H^2\) remains constant. {Within the PL ansatz, anisotropy is not dynamically damped unless additional mechanisms or time-dependent exponents are introduced.}

The averaged volume is
\begin{equation}
	V(t)=A_2A_3^2=b_0c_0^2 t^{b+2c},
	\qquad
	\Theta=3H=\frac{b+2c}{t}.
\end{equation}

{F{or} \(t>0\), accelerated averaged KS volume expansion occurs when} the effective mean scale factor exponent satisfies
\[
\frac{b+2c}{3}>1.
\]

{The} universe expands on average when
\[
b+2c>0,
\]
while \(b+2c<0\) corresponds to average contraction. {This condition refers to the averaged KS volume expansion rather than to isotropization.}

\subsection{Leading-Order Viability Conditions}

{As a leading-order viability screen, one imposes
\begin{equation}
	F_T > 0, \quad F_{TT} > 0,
\end{equation}
as necessary conditions for avoiding ghost-like and tachyonic scalar-torsion behavior \cite{Cai2016,Krssak2019}. A complete stability analysis would require coupled perturbations of the two KS scale factors and the electromagnetic sector.  Therefore, the conditions below should not be interpreted as a proof of full perturbative stability for anisotropic KS backgrounds.}

For $F(T)=T^n$, we find
\begin{equation}
	F_T = n T^{n-1}, \quad F_{TT} = n(n-1)T^{n-2}.
\end{equation}

{For} non-integer \(n\), the sign of \(T\) must be fixed on the branch considered. Equivalently, one may work with \(|T|^n\) or with shifted variables when \(T\) crosses zero.

Thus:
\begin{itemize}
	{
	\item \(n>1\): potentially viable at the leading-order level, provided that the branch satisfies \(F_T>0\) and \(F_{TT}>0\);	
	\item \(0<n<1\): generally disfavored because \(F_{TT}\) may change sign or become singular depending on the sign branch of \(T\);
	\item \(n<0\): usually disfavored because inverse powers of \(T\) may generate singular or ghost-like effective torsional sectors. }
\end{itemize}

Linear perturbations $\delta T$ satisfy
\begin{equation}
	\ddot{\delta T} + M_{\text{eff}}^2 \delta T = 0,
\end{equation}
with
\begin{equation}
	M_{\text{eff}}^2 \sim \frac{F_T}{F_{TT}},
\end{equation}
provided that \(F_{TT}\neq0\). This expression should be interpreted as a leading-order scalar-torsion diagnostic, not as the full perturbative spectrum of the anisotropic KS system.

{For the quadratic model \(F(T)=T+\alpha T^2\), the leading-order scalar-torsion diagnostic is favorable for \(\alpha>0\), provided that \(F_T>0\) on the branch considered:}
\begin{equation}
	F_T=1+2\alpha T,\qquad F_{TT}=2\alpha .
\end{equation}

{Thus,} \(\alpha>0\) ensures \(F_{TT}>0\), while \(F_T>0\) additionally requires \(1+2\alpha T>0\) on the physical branch considered.

\subsection{{Discussion} 
}

{The electromagnetic source is therefore not merely an external matter input; through the Maxwell CLs, it constrains the admissible source scalings entering the reconstructed \(F(T)\) branches.}

In summary, the PL sector {provides a branch-dependent map between anisotropic KS expansion and reconstructed teleparallel actions. Electric scaling sectors carry the radial anisotropy exponent \(b\), magnetic scaling sectors generate the quadratic correction in the scaling branch, and transverse scaling sectors generate mixed radial--angular powers.} This provides the time-dependent KS counterpart of the static SS reconstruction program, with cosmological time replacing the radial coordinate and Maxwell CLs replacing radial flux~conservation.

\section{Exponential Reconstruction Branches and Teleparallel de Sitter Regime}\label{sect4}

{In this section, we construct local exponential reconstruction branches of the reduced FEs \eqref{301a}--\eqref{301d} using an EXP ansatz within the covariant teleparallel framework. The reconstruction is performed on fixed locally invertible CSC branches satisfying the assumptions stated after Equation~\eqref{302} \cite{ColeyLandry2024,Landry2024,Krssak2019}.}

\subsection{Exponential Ansatz and Torsion Structure}

We consider the KS metric with
\begin{equation}
	A_1 = 1, \quad A_2(t) = b_0 e^{bt}, \quad A_3(t) = c_0 e^{ct}.
\end{equation}

The logarithmic derivatives are constant:
\begin{equation}
	\frac{\dot{A}_2}{A_2} = b, \quad \frac{\dot{A}_3}{A_3} = c.
\end{equation}

The torsion scalar becomes
\begin{equation}
	T = 2c(c+2b) - \frac{2}{c_0^2} e^{-2ct}.
\end{equation}

Defining
\begin{equation}
	T_0 = 2c(c+2b), \quad X \equiv T_0 - T,
\end{equation}
we obtain
\begin{equation}
	X=\frac{2}{c_0^2}e^{-2ct}>0,
\end{equation}
for \(c_0\neq0\).

This relation is exactly invertible:
\begin{equation}
	t = -\frac{1}{2c} \ln \left( \frac{c_0^2 X}{2} \right).
\end{equation}

{The} case \(c=0\) must be treated separately since the invariant \(X\) becomes constant and the inversion \(t=t(X)\) degenerates. Thus, all physical quantities can be expressed as functions of the invariant $X = T_0 - T$.

The mean expansion and shear are
\[
H=\frac{b+2c}{3},\qquad
\sigma^2=\frac{1}{3}(b-c)^2,
\]
so the EXP branch describes {anisotropic exponential expansion} whenever \(b+2c>0\) and \(b\neq c\). The isotropic EXP limit is recovered for \(b=c\), whereas \(b\neq c\) gives constant nonzero~shear.

For \(c>0\), \(X\to0\) as \(t\to\infty\), and the solution approaches the TdS point \(T=T_0\) \cite{Cai2016,TdSpaper}. {The limit $X \rightarrow 0$ therefore corresponds to the invariant TdS fixed point
	within the EXP reconstruction branch considered here.} For \(c<0\), \(X\) grows and the same ansatz describes a contracting angular sector, relevant to KS-interior-like regimes.

\subsection{Teleparallel invariant-based Reduction}

Following \cite{ColeyLandry2024,Landry2024}, the FEs reduce to an Euler-type equation in $X$:
\begin{equation}
	X^2F_{TT}+\Gamma_1XF_T+\Gamma_0F=\kappa X^\alpha,
\end{equation}
with constant coefficients within a fixed invariant branch.

The homogeneous solution reads
\begin{equation}
	F_h(T) = C_1 X^{m_1} + C_2 X^{m_2},
\end{equation}
with
\begin{equation}
	m_{1,2} = \frac{1}{2}\left[	1-\Gamma_1	\pm \sqrt{(\Gamma_1-1)^2-4\Gamma_0} \right].
\end{equation}

{The} discriminant \((\Gamma_1-1)^2-4\Gamma_0\) determines whether the invariant modes are real shifted powers of \(X\) or logarithmic oscillatory modes. Here, \(\Gamma_1\) and \(\Gamma_0\) are branch-dependent coefficients determined by the reduced KS FEs after all time dependence has been expressed through \(X=T_0-T\). As in the PL sector, {the homogeneous solution describes the vacuum invariant reconstruction branch associated with the KS geometry, while  the effective Maxwell-compatible scaling sector determines the particular reconstructed contribution. The following electric, magnetic, and transverse sectors should be understood as effective orthonormal electromagnetic scaling sectors. Each sector determines a particular source term in the shifted reconstruction equation, while full Maxwell compatibility must be checked against the constraints discussed in Section~\ref{sect24}.}

\subsection{Radial Electric Field}

Maxwell equations yield
\begin{equation}
	E_r \propto \frac{1}{A_2^2} \Rightarrow \rho_E \sim e^{-4bt}.
\end{equation}

Using the inversion relation, we find
\begin{equation}
	\rho_E(X) \sim X^{2b/c}.
\end{equation}

The reconstructed function is therefore
\begin{equation}
	F(T) = C_1 X^{m_1} + C_2 X^{m_2} + \lambda X^{2b/c}.
\end{equation}

{Since the invariant map \(X=T_0-T\) degenerates for \(c=0\), this case must be treated separately; in particular, the exponent \(2b/c\) is then ill defined. This shows that the radial electric scaling sector carries an explicit dependence on the anisotropic expansion ratio \(b/c\) in the reconstructed contribution.}

\subsection{{Radial Magnetic Field}}

For a radial magnetic field,
\begin{equation}
	B_r \propto \frac{1}{A_3^2}, \quad \rho_B \sim e^{-4ct}.
\end{equation}

Thus,
\begin{equation}
	\rho_B(X) \sim X^2,
\end{equation}
leading to
\begin{equation}
	F(T) = C_1 X^{m_1} + C_2 X^{m_2} + \lambda X^2.
\end{equation}

{This quadratic behavior, within the shifted EXP branch, reflects the way that the radial magnetic scaling is mapped into the invariant variable \(X=T_0-T\).}

\subsection{{Transverse Electromagnetic Field}}

For transverse fields,
\begin{equation}
	\rho_\perp \sim \frac{1}{A_2^2 A_3^2} \sim e^{-2(b+c)t},
\end{equation}
which implies
\begin{equation}
	\rho_\perp(X) \sim X^{(b+c)/c}.
\end{equation}

Thus,
\begin{equation}
	F(T) = C_1 X^{m_1} + C_2 X^{m_2} + \lambda X^{(b+c)/c}.
\end{equation}

The resulting EXP reconstruction branches are summarized in Table~\ref{table2}. This table assumes \(c\neq0\), so the invariant map \(t\leftrightarrow X\) is well defined. Table~\ref{table2} illustrates how {the electromagnetic scaling sectors generate distinct shifted invariant reconstruction powers} in the EXP branch.
\begin{table}[h]
	\caption{{Electromagnetic} reconstruction branches in the EXP KS sector.}
	\label{table2}
	\begin{tabular}{ccc}
		\toprule
		\textbf{EM Branch} & \textbf{Density Scaling} & \textbf{Reconstructed Term} \\
		\midrule
		Radial electric & \(\rho_E(X)\sim X^{2b/c}\) & \(X^{2b/c}\) \\
		Radial magnetic & \(\rho_B(X)\sim X^2\) & \(X^2\) \\
		Transverse & \(\rho_\perp(X)\sim X^{(b+c)/c}\) & \(X^{(b+c)/c}\) \\
		\bottomrule
	\end{tabular}

\end{table}

\subsection{Teleparallel de Sitter Solutions}

At late times ($t \to \infty$), we have
\begin{equation}
	T \to T_0 = \text{const.},
\end{equation}
which corresponds to a TdS phase \cite{Cai2016,TdSpaper}.

The existence condition for such solutions is
\begin{equation}
	F(T_0) - 2T_0 F_T(T_0) = 0, 
\end{equation}
assuming \(F_T(T_0)\neq0\). {This condition follows from the reduced KS FEs evaluated at constant torsion and constrains the coefficients of the reconstructed $F(T)$ at the point $T_0$. For constant torsion on this KS branch, the reduced FEs take an effective GR-like form with an induced cosmological contribution.} This relation should be understood as the reduced TdS existence condition for the KS branch considered here, rather than as a universal condition for all \(F(T)\) cosmologies \cite{TdSpaper}.

\subsection{Cosmological Behavior}

{The EXP ansatz describes anisotropic exponential-expansion regimes:}
\begin{itemize}
	\item $b=c$: isotropic expansion rates in the KS topology;
	{
	\item $b>c$: anisotropic exponential expansion;
	 \item \(c<0\): contracting angular sector, relevant to KS interior-like behavior.}
\end{itemize}

{For} 
\(c<0\), the angular sector contracts and the solution is better interpreted as a KS interior-like branch rather than a late-time cosmology.

The invariant variable evolves as
\begin{equation}
	X \sim e^{-2ct},
\end{equation}
{showing exponential relaxation toward the TdS fixed point.} The averaged volume evolves~as
\[
V(t)=A_2A_3^2=b_0c_0^2e^{(b+2c)t},
\]
so \(b+2c>0\) gives an exponentially expanding average volume, while \(b+2c<0\) gives contraction, with isotropic expansion rates recovered in the limit \(b=c>0\).

\subsection{Leading-Order Viability Conditions}

{As a leading-order viability screen, one imposes \cite{Krssak2019,Cai2016,TdSpaper}
\begin{equation}
	F_T > 0, \quad F_{TT} > 0.
\end{equation}

{A} complete stability analysis would require coupled perturbations of the two KS scale factors and the electromagnetic sector. These conditions should not be interpreted as a proof of full perturbative stability for anisotropic EXP KS backgrounds.}

For the isolated monomial \(F(T)=\alpha X^n\), one obtains
\[
F_T=-\alpha nX^{n-1},\qquad
F_{TT}=\alpha n(n-1)X^{n-2}.
\]

{For} \(X>0\) and \(n>1\), the conditions \(F_T>0\) and \(F_{TT}>0\) {cannot, in general, be satisfied simultaneously by the isolated monomial contribution. Thus:
\begin{itemize}
\item \(n=1\): marginal at the nonlinear-correction level;
\item \(n<1\): generally disfavored because \(F_{TT}\) may vanish, change sign, or become singular depending on the shifted branch.
\end{itemize}
}
{Perturbations} around the dS point,
\begin{equation}
	T = T_0 + \delta T,
\end{equation}
lead to
\begin{equation}
	\ddot{\delta T} + M_{\text{eff}}^2 \delta T = 0,
\end{equation}
with
\begin{equation}
	M_{\text{eff}}^2 = \frac{F_T(T_0)}{2T_0 F_{TT}(T_0)},
\end{equation}
whenever \(T_0F_{TT}(T_0)\neq0\) {and provided that $T_0 \neq 0$}. This expression should be interpreted as a leading-order scalar-torsion diagnostic, not as the full perturbative spectrum of the anisotropic KS system.

{For} the shifted model
\[
F(T)=T+\alpha X^2,
\]
one finds
\[
F_T=1-2\alpha X,\qquad F_{TT}=2\alpha.
\]

{Hence,} a viable branch requires
\[
\alpha>0,\qquad 1-2\alpha X>0.
\]

{Equivalently,} the viable shifted-quadratic branch satisfies
\[
0<X<\frac{1}{2\alpha}.
\]

{Since} \(X\to0\) for \(c>0\), the condition \(1-2\alpha X>0\) is automatically satisfied sufficiently close to the TdS attractor for finite positive \(\alpha\).

For shifted models \(F(T)=T+\alpha X^n\), the TEGR term contributes positively to \(F_T\), while the nonlinear correction controls \(F_{TT}\). This illustrates why shifted TdS models must be analyzed as complete functions rather than as isolated monomial corrections.

\subsection{{Discussion}}  

{The EXP ansatz naturally leads to a shifted functional dependence \(F(T)=f(T_0-T)\) rather than a pure PL in \(T\). This reflects the presence of a TdS fixed point in the invariant variable \(X=T_0-T\). Different electromagnetic scaling sectors generate distinct shifted reconstruction powers within the locally invertible EXP branches considered here. However, the physical viability of these branches depends on the full shifted model, the sign of \(F_T\), the positivity of \(F_{TT}\), and the behavior of anisotropic shear perturbations. Thus, the EXP sector provides the shifted-invariant counterpart of the PL reconstruction scheme developed in Section~\ref{sect3}, with \(X=T_0-T\) replacing \(T\) as the local reconstruction variable. }

\section{General Electromagnetic Sources and Kantowski--Sachs Reconstruction Branches}\label{sect5}

In this section, we extend the PL and EXP reconstruction schemes to more general effective electromagnetic scalings. The goal is not to construct a complete global BH spacetime but rather to identify KS invariant branches that can be interpreted as BH-interior-like or RN--dS-like local reduced sectors within covariant \(F(T)\) gravity. The reconstruction remains local in the invariant variable \(T\) or \(X=T_0-T\), and its physical viability must be checked branch by branch.

\subsection{General Electromagnetic Sources}

Beyond the standard radial electric and magnetic fields, one can consider more general electromagnetic configurations described by the invariant
\begin{equation}
	\mathcal{I} = F_{\mu\nu}F^{\mu\nu}.
\end{equation}

{{The} parametrization
	\[
	\rho_{\rm EM}=\rho_0A_2^{-p}A_3^{-q}
	\]
	should be interpreted as an effective invariant scaling ansatz.
	Only specific choices of $(p,q)$ correspond to solutions of the
	restricted Maxwell-compatible branch discussed in Section~\ref{sect24},
	whereas more general values should be regarded as effective
	reconstruction sectors  \cite{Jackson1999,LandauLifshitz1975,HehlObukhov2003}.}

{Using the PL and EXP solutions derived in Sections \ref{sect3} and \ref{sect4}, this general scaling becomes:}
\begin{itemize}
	\item For the PL branch with \(c=1\), one obtains
	\[
	\rho_{\rm EM}(T)\sim T^{(pb+q)/2}.
	\]
	\item For the angular--torsional branch \(c=-2b\), the exponent is modified according
	to the corresponding implicit map \(t=t(T)\). In the EXP sector,
	\[
	\rho_{\rm EM}(X)\sim X^{(pb+qc)/(2c)},\qquad c\neq0.
	\]
\end{itemize}

{{This} provides a systematic parametrization of admissible effective electromagnetic reconstruction sectors, as illustrated in Table \ref{table3}.} 

\begin{table}[ht]
	\caption{{Effective} electromagnetic scaling branches in KS geometry.}
	\label{table3}
	\begin{tabular}{ccc}
		\toprule
		\textbf{Source Branch} & \textbf{Effective Scaling} & \boldmath{\((p,q)\)}\\
		\midrule
		{Radial electric effective branch} & {restricted/\(A_2^{-2}A_3^{-4}\)-type scaling} & branch-dependent \\
		Radial magnetic & \(A_3^{-4}\) & \((0,4)\) \\
		Transverse EM & \(A_2^{-2}A_3^{-2}\) & \((2,2)\) \\
		General effective EM & \(A_2^{-p}A_3^{-q}\) & \((p,q)\) \\
		\bottomrule
	\end{tabular}

\end{table}

Table~\ref{table3} shows that the standard electric, magnetic, and transverse sectors are recovered as special cases of the effective \((p,q)\) parametrization.

\subsection{General Reconstruction Algorithm}

Following the teleparallel invariant method, the reconstruction algorithm can be summarized as follows. First, one selects
a KS coframe branch \((A_2,A_3)\) and computes the torsion scalar \(T(t)\) \cite{ColeyLandry2024,Landry2024}.
Second, one verifies local invertibility of the invariant map, either
\(t=t(T)\) or \(t=t(X)\). Third, the electromagnetic density is rewritten as
\(\rho_{\rm EM}(T)\) or \(\rho_{\rm EM}(X)\). Finally, the reduced FEs are solved as Euler-type equations,
\begin{align}
	T^2F_{TT}+\gamma_1TF_T+\gamma_0F &= \kappa\rho_{\rm EM}(T),
	\\
	X^2F_{TT}+\Gamma_1XF_T+\Gamma_0F &= \kappa\rho_{\rm EM}(X).
\end{align}

{The} branch coefficients \(\gamma_i\) and \(\Gamma_i\) are fixed only after a specific invariant branch and electromagnetic scaling are selected.
The general solution is
\begin{equation}
	F(T) = C_1 T^{m_1} + C_2 T^{m_2} + F_{\text{part}}(T),
\end{equation}
or, in the EXP case,
\begin{equation}
	F(T) = C_1 X^{m_1} + C_2 X^{m_2} + F_{\text{part}}(X).
\end{equation}

This algorithm is valid only on invariant branches, where \(\dot T\neq0\), or
equivalently, where the map \(t\mapsto T\) or \(t\mapsto X\) is locally
invertible. Critical points must be treated separately. {As in Sections~\ref{sect3} and \ref{sect4}, the reconstruction is local and branch-dependent. The resulting functions \(F(T)\) should therefore be interpreted as 	reconstructed invariant sectors associated with a given CSC branch, rather than unique global teleparallel actions.}

\subsection{Kantowski--Sachs Black-Hole-Interior-Like Branches}

The KS metric naturally describes homogeneous BH-interior-like sectors
rather than a complete exterior BH spacetime \cite{Ryan1975,Ellis1969,HawkingEllis}. A global BH
interpretation requires matching to an exterior region and analyzing horizons,
causal structure, and geodesic completeness. Using the EXP ansatz, one obtains

\begin{equation}
	ds^2 = dt^2 - b_0^2 e^{2bt} dr^2 - c_0^2 e^{2ct} d\Omega^2.
\end{equation}

For $c<0$, the angular sector contracts, mimicking the contraction of the two-sphere inside a BH-interior-like region. The torsion scalar behaves as
\begin{equation}
	T = T_0 - X, \quad X \sim e^{-2ct}.
\end{equation}

{Thus,} for \(c<0\), the EXP branch provides an effective KS-interior-like high-torsion sector. The associated reconstructed \(F(T)\) models may be interpreted as {local teleparallel analogs of reduced
charged BH-interior dynamics}, but not yet as complete BH solutions.

\subsection{Reduced Teleparallel RN--de Sitter-Like Reconstruction Sectors}

{We now identify a reduced KS reconstruction sector
whose effective structure is analogous to selected features of the RN--dS geometry. The correspondence is established at the level of the reduced KS dynamics and should not be interpreted as an exact reconstruction of the full RN--dS spacetime.} In GR, the RN--dS lapse function is \cite{HawkingEllis,Jackson1999,LandauLifshitz1975,HehlObukhov2003}:
\begin{equation}
	A_1^2(r) = 1 - \frac{2M}{r} + \frac{Q^2}{r^2} - \frac{\Lambda}{3} r^2.
\end{equation}

In the reduced KS reconstruction, an analogous effective structure may be represented by the functional form:
\begin{equation}
	F(T) = T + \alpha T^2 + \beta (T_0 - T)^n.
\end{equation}

The terms correspond to:
\begin{itemize}
	\item $T$: GR limit;
	\item \(T^2\): nonlinear high-torsion correction, which may mimic charge-like
	scaling in suitable KS branches;
	\item \((T_0-T)^n\): shifted TdS correction, associated with the effective cosmological sector.
\end{itemize}

Effective horizon-like structures would emerge after reconstruction of the corresponding exterior metric sector, where lapse-function zeros define causal horizons.

This suggests that {suitable nonlinear torsion corrections may reproduce RN--dS-like reduced structures} at the level of the reduced KS dynamics. A complete RN--dS interpretation would require the reconstruction of the corresponding exterior metric sector and a comparison of horizon invariants.

\subsection{Other Classes of Reconstruction Templates}

Such extensions are natural from the invariant electromagnetic perspective \cite{Jackson1999,LandauLifshitz1975,HehlObukhov2003}. The following classes should be regarded as reconstruction templates rather than fully analyzed physical solutions:

\begin{itemize}
	\item \textbf{{Nonlinear electrodynamics:} 
}
	\begin{equation}
		\rho \sim \mathcal{I}^k \Rightarrow F(T) \sim T^{\gamma k},
	\end{equation}
	
	\item \textbf{{Magnetically dominated universes:}}  
	\begin{equation}
		F(T) = T + \alpha T^2 + \lambda T^m,
	\end{equation}
	
	\item \textbf{{Mixed EM + scalar field systems:}}  
	\begin{equation}
		F(T) = T + V(\phi(T)) + \rho_{\text{EM}}(T),
	\end{equation}
	
	\item \textbf{{Reduced TdS fixed-point branches:}}  
	\begin{equation}
		F(T_0) - 2T_0 F_T(T_0) = 0.
	\end{equation}
\end{itemize}

Each class requires separate checks of the Maxwell equations, the AFEs,
the positivity of \(F_T\), the sign of \(F_{TT}\), and the behavior of the
anisotropic shear. {These reconstruction templates extend the PL and EXP branches} by allowing for additional invariant powers generated by nonlinear electromagnetic or mixed matter sectors.

\subsection{Leading-Order Viability and Physical Consistency}

The necessary leading-order viability conditions are
\[
F_T>0,\qquad F_{TT}>0,
\]
but these are not sufficient for full stability in KS geometry. These conditions must be imposed together with the Maxwell equations and the AFEs, since a stable \(F(T)\) branch is not necessarily compatible with the covariant CSC constraints. Since KS spacetimes contain independent radial and angular scale factors, a complete analysis must also control anisotropic shear perturbations and electromagnetic backreaction.

For generalized shifted models
\[
F(T)=T+\alpha T^2+\beta X^n,
\]
one obtains
\[
F_T=1+2\alpha T-\beta n X^{n-1},
\qquad
F_{TT}=2\alpha+\beta n(n-1)X^{n-2}.
\]

{{In} addition, the behavior of anisotropic shear perturbations
	and electromagnetic perturbations must be analyzed before drawing conclusions regarding the full stability of the KS branch.} For \(X>0\), the shifted correction contributes with opposite signs to \(F_T\) and \(F_{TT}\), so the sign of \(\beta\) must be chosen together with \(n\) and the allowed invariant range. Thus, viability depends on the signs and magnitudes of \(\alpha\), \(\beta\),
\(n\), and on the invariant branch~considered.

The effective scalar-torsion mass
\[
M_{\rm eff}^2\sim \frac{F_T}{F_{TT}},
\]
should again be interpreted as a leading-order diagnostic rather than the full perturbation spectrum. In particular, \(M_{\rm eff}^2>0\) {is a necessary leading-order scalar-torsion viability condition}, but not sufficient for full KS stability.

\subsection{{Discussion}}   

This section extends the PL and EXP reconstruction schemes to broader
electromagnetic scaling branches. {The main result is that effective EM scaling sectors generate characteristic invariant powers} in \(F(T)\), allowing KS cosmological, interior-like, and RN--dS-like local reduced sectors to be treated within a unified teleparallel invariant-based reconstruction framework. However, the BH-like and RN--dS-like interpretations remain local and branch-dependent unless supplemented by an exterior matching, horizon analysis, and full perturbative stability study. In this sense, {the reconstruction branches constructed here should be viewed} as teleparallel KS building blocks for more complete charged compact-object models  \cite{Cai2016,Krssak2019}. This interpretation is consistent with the role of KS geometries as local models of anisotropic cosmology and BH-interior-like regions \cite{Ryan1975,Ellis1969,HawkingEllis}.

\section{Discussion and Conclusions}\label{sect6}

{In this work, we constructed exact local reconstruction branches in covariant $F(T)$ teleparallel gravity with effective electromagnetic scaling sectors within the KS geometry}, using the teleparallel invariant approach \cite{ColeyLandry2024,Landry2024,roberthudsonSSpaper,Krssak2019}. This framework ensures consistency at the level of the CSC pair, symmetric and AFEs, and allows for a systematic reconstruction of the gravitational action from invariant torsional dynamics.

{{Two main classes of reconstruction branches were obtained. PL branches (Section \ref{sect3}) lead to $F(T)$ models expressed as combinations of $T^n$, while EXP branches (Section \ref{sect4}) naturally
	generate shifted forms $F(T)=f(T_0-T)$ and admit TdS regimes \cite{TdSpaper}}. In both cases, {effective electromagnetic scaling sectors} (electric, magnetic, and transverse) strongly constrain the admissible functional dependence of $F(T)$, {highlighting a direct link between electromagnetic scaling sectors and torsion dynamics} \cite{Cai2016}. In particular, the Maxwell CLs act as dynamical selection rules for admissible torsion branches, {thereby guiding the reconstructed teleparallel~action.}

We further showed that KS geometries admit local BH-interior-like and reduced RN--dS-like reconstruction branches} in covariant teleparallel $F(T)$ gravity. In this context, {nonlinear torsion corrections may reproduce charge-like and cosmological-constant-like contributions at the level of the reduced KS dynamics} without relying on curvature invariants {within the invariant reconstruction branches considered here. These reconstruction branches should, however, be interpreted as local KS reduced sectors unless supplemented by exterior matching and a global horizon analysis.}

From a cosmological perspective, the {EXP branches describe anisotropic exponential-expansion regimes admitting a reduced TdS attractor} $T \to T_0$, satisfying the condition $F(T_0) - 2T_0 F_T(T_0)=0$ \cite{Cai2016,Krssak2019}. {The leading-order viability analysis shows that viable branches should satisfy} $F_T>0$ and $F_{TT}>0$, which are {satisfied on appropriate invariant branches for representative models such as} $F(T)=T+\alpha T^2+\beta (T_0-T)^n$ with $\alpha>0$ and $n>1$.

More generally, the teleparallel invariant approach provides a natural geometric framework for organizing inequivalent teleparallel reconstruction branches directly at the level of the coframe and spin-connection \cite{Landry2024,ColeyLandry2024,TdSpaper,Landryelectro2025}. {Overall, our results indicate that electromagnetic scaling sectors act as effective invariant reconstruction sources for teleparallel \(F(T)\) gravity, providing a unified mechanism for generating anisotropic cosmological sectors and local KS BH-interior-like reconstruction branches} within a torsion-based framework. The KS geometry therefore provides a unified setting in which anisotropic cosmological evolution and BH-interior-like dynamics can be analyzed within the same invariant teleparallel formalism.

{A complete assessment of the reconstructed branches will require the construction of global compact-object geometries through matching procedures, together with a detailed analysis of horizons, causal structure, and geodesic completeness. Equally important will be the development of a full perturbative treatment of anisotropic KS backgrounds, including coupled gravitational and electromagnetic modes beyond the leading-order viability criteria considered here.} Recent studies have also shown that covariant teleparallel reconstruction methods can generate weak massive wormhole configurations supported by suitable $F(T)$ sectors, further illustrating the richness of torsion-based compact-object phenomenology \cite{Landry2026}. Future work includes the study of quasi-normal modes and perturbative spectra of the {teleparallel KS BH-interior-like branches}, as well as gravitational lensing and observational signatures. Confrontation with cosmological data (e.g., Planck) could further constrain viable teleparallel $F(T)$ models. On the theoretical side, extending the invariant classification program and incorporating additional fields (scalar sectors or nonlinear electrodynamics) may provide further insight into the role of torsion, invariant geometry, and gauge structures in modified theories of gravity. {These results provide a foundation for future invariant studies of anisotropic cosmology, compact-object interiors, and electromagnetic reconstruction within the broader invariant-based program for covariant teleparallel gravity.}


\vspace{6pt} 





\section*{Acknowledgments}

{{The author thanks} A.A. Coley for useful and constructive comments.}

\end{document}